# A Systematic Review of Tracing Solutions in Software Product Lines

Zineb Mcharfi [a]*, Bouchra El Asri [b], Abdelaziz Kriouile [c]

[a,b,c]*IMS Team, SIME Laboratory, ENSIAS, Mohammed V University, Rabat, Morocco*
[a]*Email: zineb.mcharfi@gmail.com*
[b]*Email: elasri_b@yahoo.fr*
[c]*Email: kriouile@ensias.ma*

**Abstract**

Software Product Lines are large-scale, multi-unit systems that enable massive, customized production. They consist of a base of reusable artifacts and points of variation that provide the system with flexibility, allowing generating customized products. However, maintaining a system with such complexity and flexibility could be error prone and time consuming. Indeed, any modification (addition, deletion or update) at the level of a product or an artifact would impact other elements. It would therefore be interesting to adopt an efficient and organized traceability solution to maintain the Software Product Line. Still, traceability is not systematically implemented. It is usually set up for specific constraints (e.g. certification requirements), but abandoned in other situations. In order to draw a picture of the actual conditions of traceability solutions in Software Product Lines context, we decided to address a literature review. This review as well as its findings is detailed in the present article.

*Keywords:* Traceability; Software Product Line; Systematic Review

1. Introduction

Traceability can be defined as the ability to describe and follow the life of an artifact in forward and backward directions [1]. It deals with the interconnections between components [2], the creation and maintenance of consistent documentation, the satisfaction of all specifications [3] as well as the independence from individual knowledge [4].

Traceability helps to improve the quality of the system. It makes possible to choose the architecture that meets the needs and constraints of the project, to identify errors and to facilitate communication between the various stakeholders of the project [5]. It provides significant support in the maintenance and evolution phase and enables the analysis and control of changes impact on system components [6], a very important point in the context of Software Product Lines.

For Software Product Lines, traceability can be useful in the development phase, for short-term needs such as validation and verification of the implementation of all functional requirements, or in maintenance, for long-term needs such as understanding how artefacts work and how they relate to their environment, as well as for driving changes by allowing impact analysis and components reuse [2], [7], [8].

However, many difficulties can be faced when adopting a traceability solution in the context of Software Product Lines [9] : larger documentation, heterogeneity of documents, the need to link between different products and between the products and the architecture….

Some works focus on the problems of traceability and variability in Software Product Lines in the relationship between artifacts [5], [10], [11]; others deal with traceability through the use of meta-models while implementing the Product Line [6], or, like in [12], a reference model is established to automate traceability.

In order to get an idea of the research work already carried out as part of traceability in Software Product Lines, we decided to conduct a systematic review of the literature. By this review, we are not interested in traceability in Software Product Lines in general, but rather in the traceability solutions proposed in literature, in what context are they proposed and in what forms are they presented (tools, frameworks ...).

The remaining of this paper is organized as follow. In section 2 we describe the methodology adopted for our systematic review. In section 3 we detail the results of the systematic review applied in the context of Software Product Lines tracing solutions and in order to answer the addressed research questions. We conclude in section 4 with a synthesis of results and findings.

## 2. Systematic Review methodology

Before starting a research work, it is important, even primordial, to know the progress of studies previously conducted in the field of interest, the results obtained, as well as the points still open to research. Conducting a literature review helps answering these questions and positioning our research work.

In order to be as exhaustive and objective as possible while conducting its review of the literature, a well-defined approach should be adopted. Therefore, as part of our research work, we adopted the process described in [13] to conduct our systematic review.

We briefly describe in the following the systematic review process according to [13], before presenting its application as part of our work.

As shown in figure 1 bellow, the review process adopted is composed of three main phases: (i) Review planning, (ii) review conduction and (iii) reporting.

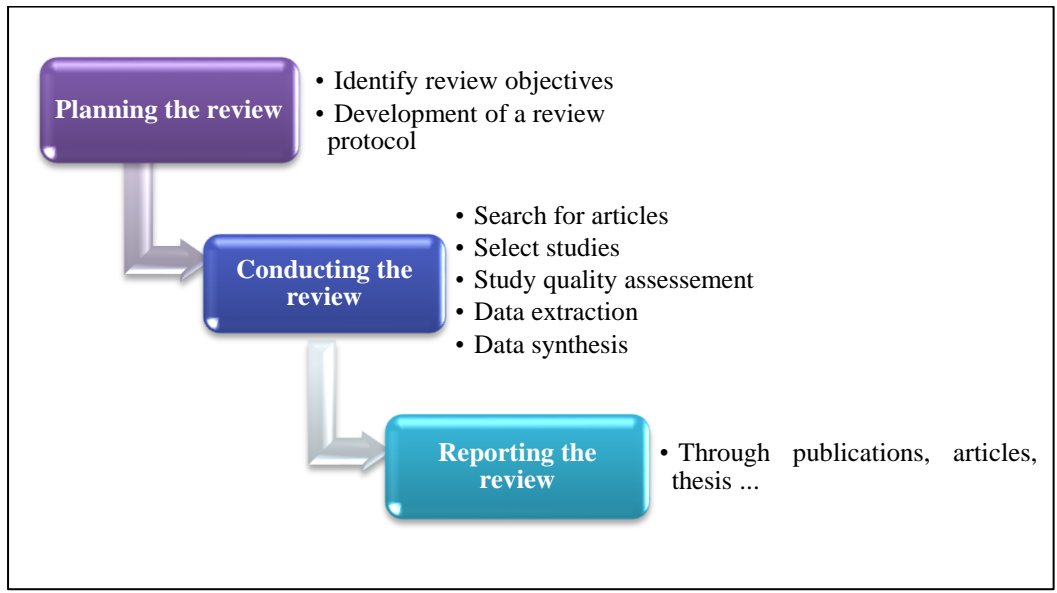

**Figure 1**: Review process according to [13]

The first phase, planning the review, consists mainly on answering the questions: what are the objectives behind conducting that review? What are the points we would like to highlight? It is also at this phase that we define a protocol to follow while conducting our literature review. This protocol defines the research questions addressed to answer review objectives, the databases where to look for articles, search expression and keyword to be used to search for articles, the criteria and filters to apply in order to select the most relevant studies, data and information to extract from retained studies in order to answer research questions, and finally the graphics and statistics to be used to analyze and synthesize extracted data.

In the second phase related to conducting the review, we apply what has been defined in the planning phase: search for articles, select studies by applying pre-defined filters, verify the quality of the selected studies, extract data from the retained articles according to a pre-defined form, and then analyze and synthesize the results.

The latest phase consists on reporting and sharing the review findings through publications, thesis…, which is the main objective of the present article.

**3. Our literature systematic review of tracing solutions in Software Product Lines**

*3.1. Review planning*

In this section of the literature review, we identify the goals behind the review and develop a review protocol that we follow to conduct our study.

*3.1.1. Review objectives*

Through this literature review, we seek to identify the progress of researches related to traceability in Software Product Lines. We mainly target the traceability solutions proposed in the literature for Software Product Lines. We want to know, through this systematic review of the literature, how this traceability is presented (using a tool, a framework, a model ...), at what stage it is applied (Domain Engineering, Applications Engineering, Maintenance), what are the dimensions covered by traceability, and does it cover them completely or partially?

All these questions are intended to make us understand and know, in addition to the overall progress of the work in the literature, the elements that can impact a given traceability solution in the Software Product Lines, so as it helps guide our subsequent contributions.

This literature review should also allow us to understand how traceability is integrated into product lines, despite the complexity of this system.

### 3.1.2. Review protocol

Below are the most important elements of our research protocol: the research questions we established, the research strategy followed, and the chosen selection criteria.

**Research questions**:

In order to reach our research objectives, we established a list of questions that we will try to answer through our systematic review

**RQ1: At which phase (Domain Engineering, Application Engineering or Maintenance) is traceability implemented?**

A Software Product Line can be divided into two main phases, or even three: (i) Domain Engineering phase during which the Product Line platform and core assets are created, and the variability defined. It is during this phase also that the plan and scope of the project are established, (ii) Application Engineering phase where the elements of Domain Engineering are instantiated and the choice of the variation scenario made, according to the product to be generated and (iii) Maintenance phase for maintaining and evolving the system.

Depending on the traceability policy adopted, trace links can be created in one of these phases. It is therefore important, in the context of our research, to know what exists in literature, i.e. the phase of creation of trace links, depending on the chosen traceability solution.

**RQ2: Which trace links between artefacts are covered by tracing process?**

Due to variability dimension in the Product Lines, traceability can be implemented at different scales: horizontally, vertically, between components from the same phase (Domain Engineering - Domain Engineering ...) or different phases (Domain Engineering - Applications Engineering ...). Therefore, it is interesting to analyze the results of our systematic review from this angle, in order to be able to deepen our research and study traceability links from the simplest to the most complex one to implement.

Thus, we used the traceability links classification defined in [5] :

Refinement: this type of traceability links defines the relationship between elements belonging to different levels of abstraction, in the same phase of the Product Line (Domain Engineering or Application Engineering). This can be for example the relationship between two models in Domain Engineering, such as features model and design model.

Similarity: Corresponds to traceability links between artifacts of the same level of abstraction, in the Domain Engineering or Application Engineering phase. For example, they may be links between class diagram and component diagram in UML, both of which are elements of the same level of abstraction (design) [5].

Variability: Two types of variability are defined in [5]: the variability between a variation point and its variants, at Domain Engineering level, called "realize" variability, and the variability between a variation point at Domain Engineering level and its instantiation at the Applications Engineering level, called "use" variability.

Versioning: This type of variability links between two versions of the same artifact. It represents traceability in time.

**RQ3: What type of traceability is it? Full, ad hoc or targeted?**

Whether to trace all the links or some, on demand or according to specific criteria, the traceability policy adopted varies and the cost of its implementation too. By addressing this question, we wish to know the trends in literature, in order to compare it with our future research works.

**RQ4: What is the proposed solution to implement traceability in the Software Product Line?**

This question aims at orienting us towards the tendencies of type of solutions (models, meta-models, frameworks ...) proposed in literature in order to answer the problem of traceability integration in Software Product Lines. We also seek to provide elements related to the chosen method (how the framework is implemented, which model is used ...).

**RQ5: Is it a manual or automatic traceability?**

There is a big difference in terms of costs between manual and automatic traceability: the first is certainly less expensive when setting up the solution, but its exploitation could be more restrictive in the context of a system as complex as the Software Product Lines. An automatic solution (by automatic we are interested in automated solutions that use a tool) could be costly, but facilitates the exploitation and the use of traceability in the context of Software Product Lines [2].

**Search strategy:**

- **Search databases**

In order to select articles to be analyzed for our research work, we have identified the following sources that we consider to be the most relevant and which bring together a significant number of research articles. These sources include libraries/digital databases and newspapers. Note that the search from these sources also includes articles published in the newspapers Journal of Systems and Software (JSS) and Information and Software Technology (IST) (source Elsevier), as well as the journal Software & Systems Modeling (source Springer link).

**Table 1**: Search databases

| Database | Link |
|---|---|
| **ACM** | https://dl.acm.org/ |
| **IEEE Xplore** | https://ieeexplore.ieee.org/ |
| **Elsevier** | https://www.elsevier.com/en-xm |
| **Springer link (Software & Systems Modeling)** | https://link.springer.com |

- **Search terms and keywords**

We used the following search expression to search for the articles:

**("software" OR "system") AND ("SPLE" OR "SPL" OR "Product Line" OR "Product Family") AND ("traceability" OR "tracing" OR "trace").**

We adapted this expression depending on the database, to be in phase with the research constraints imposed by each of the selected libraries or digital databases.

- **Study inclusion and exclusion criteria:**

To refine our search, we decided to retain only the articles that meet the criteria below:

- Research article,
- Written in English
- Published between 2008 and 2018
- Number of pages exceeding 5 pages
- Offers interesting and enriching elements to our work
- Not duplicated, in which case the article that gives the more detailed information will be kept.

- **Study quality assessment:**

In order to select only the most relevant articles, we have decided to consider, as quality criteria of the article being evaluated, the number of times it has been referenced, taking into consideration that recently published articles will still be cited very little.

- **Data extraction strategy:**

The data is extracted form retained articles according to the elements in the table below:

**Table 2**: Data extraction elements

| Element | Example |
|---|---|
| Title | Variability extraction and modeling for product variants |
| Authors | L. Linsbauer, R. E. Lopez-Herrejon, and A. Egyed |
| Abstract | … |
| Keywords | Traceability ; SPL |
| Publication Year | 2017 |
| Publication Database | Springer |
| Publication Source | Software & Systems Modeling |
| QR1 | Maintenance |
| QR2 | Refinement & Similarity |
| QR3 | Targeted |
| QR4 | Approach + Methodology |
| QR5 | Manual |

- **Synthesis of the extracted data:**

In order to allow a better analysis of the extracted data, we will present the results through the following graphs:

o graph in sector to schematize the distribution of articles selected for this systematic review according to the publication database,

o graph in sector to schematize the distribution of the articles retained for this systematic review according to the source of publication,

o evolution curve of the publications per year during the period of our study (from 2008 to 2018),

o Histogram representing the number of articles that processed traceability, distributed according to the phases of a Software Product Line (Domain Engineering, Applications Engineering, Maintenance),

o Histogram representing the number of articles by type of traceability links covered (refinement, similarity, variability, versioning)

o Histogram representing the number of articles per traceability approach adopted (full, ad hoc or targeted),

o Histogram representing the number of contributions by nature of the proposed solution (framework, model, meta-model, approach, algorithm, tool)

o Histogram representing the number of articles offering an automatic, semi-automatic or manual solution.

*3.2. Conducting the review*

This section is a direct application of what have been established in the review protocol. We search for articles according to a defined expression, we select the studies and go through several filters to get our final set of articles so as we can extract relevant data and make our analysis.

*3.2.1. Search results:*

9,449 articles were selected after the execution of predefined search keywords on the previously identified databases. Articles are distributed as follow:

**Table 3:** Search results by search database

| Database | Number of articles |
| --- | --- |
| **ACM** | 967 |
| **IEEE** | 83 |
| **Elsevier** | 8 303 |
| **Springer link** | 96 |

*3.2.2. Study selection:*

Applying the type of publication, year, number of pages, and language filters selected 1,072 items distributed as follows (Table 4):

**Table 4:** Number of articles retained after application of the filters: type of publication, year, number of pages, and language

| Database | Number of articles |
| --- | --- |
| **ACM** | 690 |
| **IEEE** | 76 |
| **Elsevier** | 219 |
| **Springer link** | 87 |

We then selected, among the remaining articles, those whose title, keywords and abstract analysis allowed us to identify relevant information in relation to our work and the defined research questions.

After this operation, 20 articles were retained (Table 5):

**Table 5:** Number of articles retained after application of the filters: title, keywords and abstract

| Database | Number of articles |
|---|---|
| ACM | 11 |
| IEEE | 4 |
| Elsevier | 2 |
| Springer link | 3 |

*3.2.3. Study quality assessment*

Applying the evaluation criteria related to quality of the work, namely the number of times that the article was referenced, depending on publication year, and in addition to the detailed reading of its contents, we retain 15 articles distributed as follows (Table 6):

**Table 6:** Number of articles retained after quality assessment

| Database | Number of articles |
|---|---|
| ACM | 7 |
| IEEE | 3 |
| Elsevier | 2 |
| Springer link | 3 |

These 15 articles, retained after application of the previously detailed criteria, represent the bases for the upcoming analysis according to the research questions we have defined. The number of these articles could be considered as relatively low, this is to the fact that through this systematic review of the literature, we seek to answer specific and targeted questions, in order to guide our research work.

*3.2.4. Data extraction*

After selecting the most relevant articles, we extracted data from each retained study according to the elements previously defined in Table 2.

This data is analyzed through several indicators as illustrated in the next paragraph.

*3.2.5. Data synthesis*

As previously announced, we reproduce the extracted data in the graphs below. We will first analyze the metadata of selected articles, namely statistics by database of extraction, by source and by year of publication. We will then analyze the content of these articles, according to the research questions previously mentioned.

We present, in table 7 below, the percentage of retained articles by database compared to the articles selected after the application of search expression to the same database.

It can be seen from this table that the percentages of the articles selected are very low, 1.2% on average. Several articles were filtered during the application of our selection criteria. First, the inclusion and exclusion criteria included the year of publication filter. We have retained only the articles published during the last ten years, between 2008 and 2018, in order to have a review of the literature recent and up to date. It therefore emerges that, during this period, the research publications related to Software Product Lines did not focus sufficiently on traceability issue. This is a general trend, traceability issues, not only for Software Product Lines but in software engineering in general, are little discussed and approached given their complexity..

The second filter applied was the one related to reading title, abstract and keywords. Several articles have also been filtered at this level. This is explained by the fact that, in these articles, traceability is only mentioned, as an interesting concept for managing variability in Software Product Lines, or as a support element in the process of setting up Product Lines. No traceability implementation solution is proposed, which is a key objective of our research and our systematic review of the literature.

**Table 7:** Percentage of retained articles by database compared to the articles selected after the application of search expression

| Database | Number of articles from the initial extraction | Nombre of retained articles | Percentage from the initial extraction |
|---|---|---|---|
| **ACM** | 967 | 7 | 1% |
| **IEEE** | 83 | 3 | 4% |
| **Elsevier** | 8 303 | 2 | 0% |
| **Springer link** | 96 | 3 | 3% |
| **Total** | **9 449** | **15** | |

Figure 2 shows the percentage of articles by database for the 15 papers selected through the systematic review. It turns out that almost 50% of the articles selected for our analysis were extracted from the ACM digital library. Indeed, the majority of these articles were communicated at the International Software Product Line Conference (SPLC), which itself contains 27% of the selected publications (figure 3), and whose articles are published in ACM. The second rank, in terms of the highest percentage, is shared by publications of IEEE (20% of the selected articles), and those of Springer, whose publications are communicated mainly in the journal Software & Systems Modeling (20% of publications selected according to the graph in figure 3). The Journal of Systems and Software takes the last place as source of publications, with 13% of articles selected (figures 2 and 3).

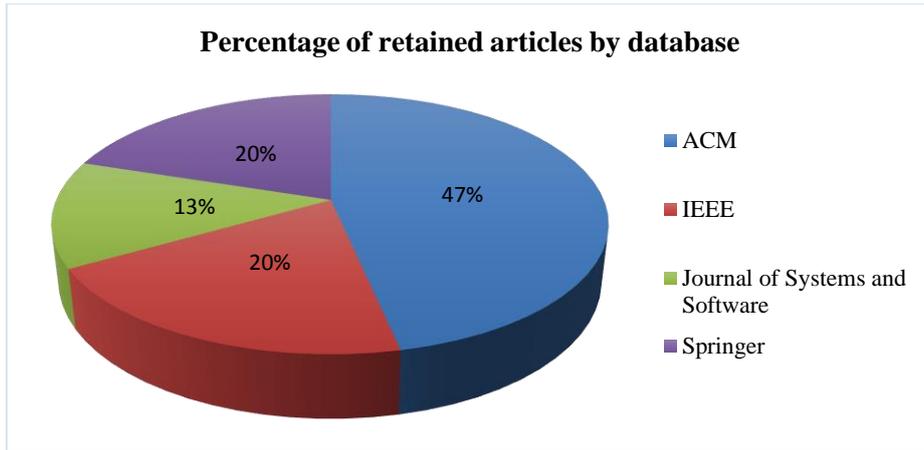

**Figure 2:** Percentage of retained articles by database

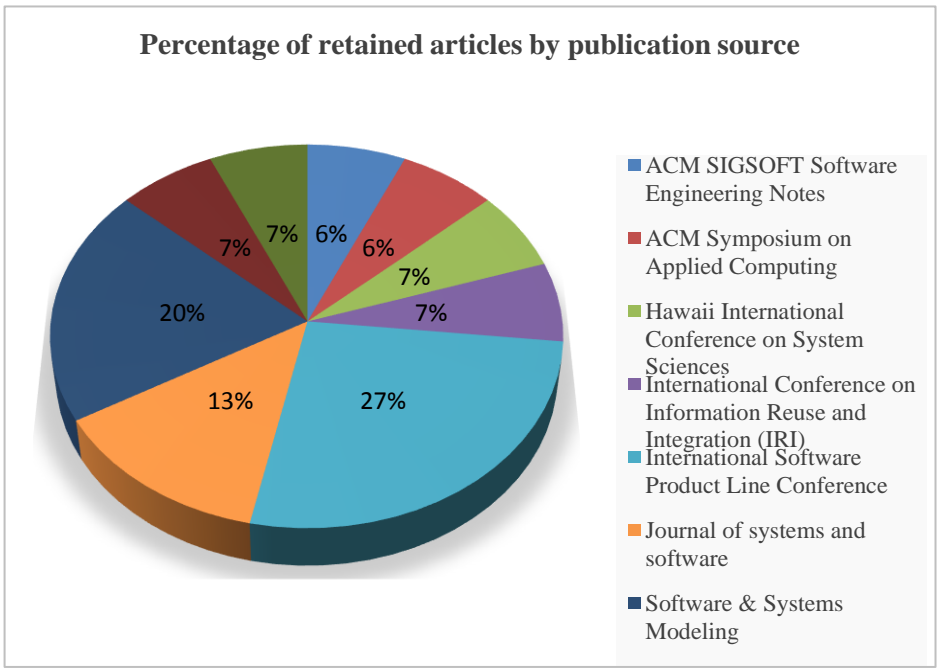

**Figure 3:** Percentage of retained articles by publication source

Regarding the distribution of articles retained by year of publication during the period selected for our systematic review (between 2008 and 2018), we notice that the number of articles is globally evolving at the same rate, with an average of 2 articles per year. We can therefore deduce that the subject of traceability solutions in Software Product Lines has received the same level of interest over the past decade, with no notable peaks of interest (Figure 4).

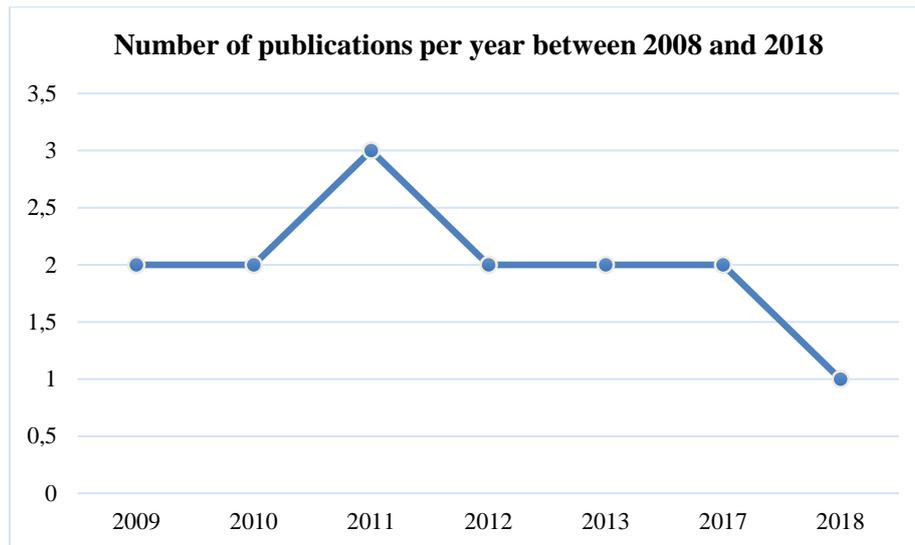

**Figure 4:** Number of publications per year between 2008 and 2018

The following statistics are related to the data collected from the selected articles, source of analysis of our systematic review.

We start with the first research question:

**RQ1: At which phase (Domain Engineering, Application Engineering or Maintenance) is traceability implemented?**

As explained earlier, there is 3 main phases in the life of a Software Product Line: Domain Engineering, Application Engineering and Maintenance. Through this analysis, we try to understand the trend of research works in terms of the implementation phase chosen.

Consider the graph figure 5. It turns out that about 66% of the articles offer solutions where traceability is implemented in the Domain Engineering phase (8 articles offering traceability only in Domain Engineering [6], [12], [14]–[19], while two of them are implementing traceability in the Domain Engineering and Application Engineering phases [5], [20]. The first group of articles, the one that focuses only on Domain Engineering, includes articles mainly offering modeling solutions, impacting the architecture of the proposed solution for the entire Software Product Line. Articles [5] and [20] also focus on the relationship between Software Product Lines and generated products. In [5], the authors use model-driven techniques to provide a framework for traceability in Software Product Lines. In this framework are treated different traceability links, especially those relating to the instantiation of variability, and therefore to the link between Domain Engineering and Application Engineering. The work in [20] describe an approach to manage the traceability in Software Product Lines through links between different models, and also between the levels Product Line and product.

Traceability solutions in Domain Engineering are therefore the predominant, followed by those in the Maintenance phase.

In fact, 5 articles, [21]–[25], which represents about 33% of articles selected, propose a traceability solution in the maintenance phase. These are mainly reengineering solutions to manage the evolution of the Product Line. In [21], the author proposes a solution for extracting traceability links from products with variation points in order to build a derivation graph from the retrieved information. The article [22] presents an approach for managing the online update of a Product Line while minimizing interruptions. In [23] is processed the re-engineering of existing products, created by simple copy and paste operations without any systematic reuse policy, into Software Product Lines. The article proposes a method for linking each piece of code from the existing end product to the corresponding feature, based on a code-topic method for creating groups of similar classes. Two types of code-topic methods are used: the textual ones to find the keywords in the comments and identifiers of the source code, and the structural ones to group the methods of call and inheritance in the source code. The article [24] proposes an algorithm to identify overlaps between the features and between the codes at the level of variation points of an already existing product, in a process of establishing the correspondence between features and source code. In [25], the objective is to find the inconsistencies of variability between the specifications and the source code of an existing product.

We do not find an article proposing traceability implementation in the Application Engineering phase alone, which is logical because at this stage the system is already in place, all what is needed is instantiating the components and choosing the desired combination through the points of variation to output a product. The implementation of traceability at this level would imply the introduction of new components, and thus a go back to the Domain Engineering phase, in which the base of the Product Line is set up.

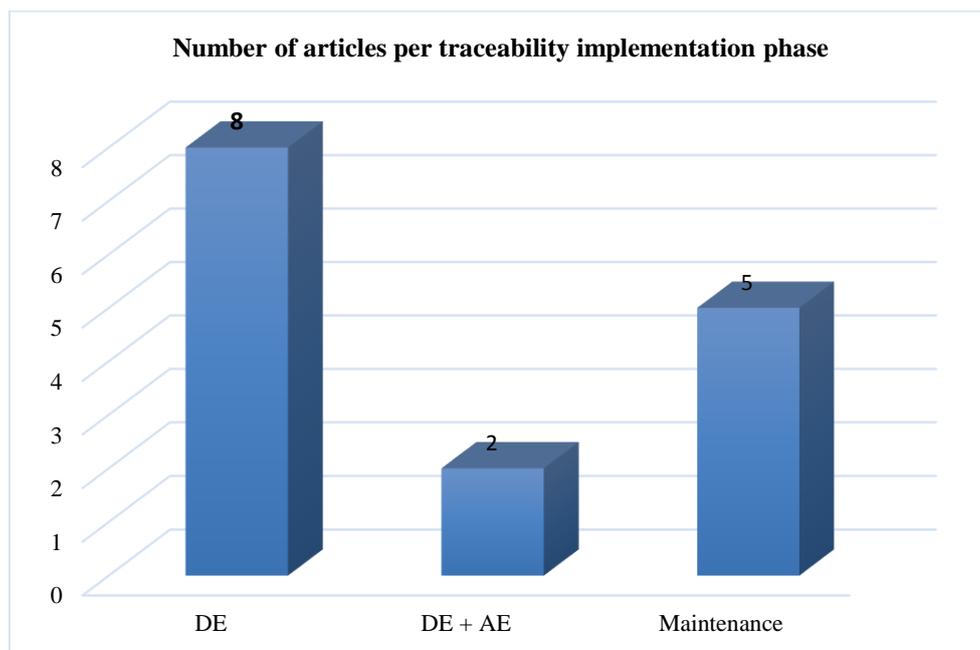

**Figure 5:** Number of articles per traceability implementation phase

**RQ2: Which trace links between artefacts are covered by tracing process?**

As previously presented, we classified the solutions proposed in the articles selected by type of proposed traceability links. To do this, we analyzed the articles, extracted the traceability links used in each of the proposed solutions, and then classified them according to the four types defined in [5]: Refinement, similarity, variability ("realize" and "use") and versioning. An article can use one or more types of links.

According to the graphic figure 6, the refinement and variability-realize trace links are the most used. This is directly related to the analysis made for the first research question from which it appears that the majority of the articles deals with traceability in the domain engineering phase. Indeed, several articles propose solutions impacting the traceability starting from features or requirements and up to the architecture or implementation phases [6], [15], [16], [18], [20], [21], [23]–[25]. Traceability of the variation points is also discussed, especially by the articles [6], [14]–[16], [19], [20], [23]–[25], which corresponds to a traceability of type variability-realize. Versioning represents the least discussed type (the two subtypes, realize and use, can be grouped for variability traceability), with only three articles [5], [6], [22]. These findings inform on tendency of research that is focused on modeling and architecture solutions, rather than versioning, even though the latter remains as important to the integrity of the proposed solution.

The research question to follow addresses the issue of type of traceability.

**RQ3: What type of traceability is it? Full, ad hoc or targeted?**

Depending on the proposed solution and the traceability links it impacts, the articles can be classified by type of traceability: full, targeted or adhoc. Figure 7 presents the result of the articles' classification by type of traceability.

The first finding we can make following the observation of this graph is that adhoc traceability is not addressed. This would be due to the fact that, the objective of our Systematic Review is to analyze solutions (frameworks, models, ...) of traceability in Software Product Lines, any work proposing an unstructured approach is then rejected through the filters of review.

Another finding: The two approaches, full and targeted, are quite equivalent in literature, with a slight advantage for targeted traceability. Some articles offer automated solutions that make it possible to create whether all the links or some of them, depending on the objectives, types of traces desired ... This is the case of articles [5] and [12]. Articles offering full traceability [6], [15]–[18] focus on end-to-end traceability solutions, from features to their according implementation. They are also part of the articles presenting links of traceability type refinement, except the article [17] which proposes traceability links of type variability-use between the features in phase of Engineering Domains and those in the Application Engineering phase.

For articles offering targeted traceability, they mainly deal with solutions that are directly related to variability points. This is particularly the case for articles [14], [19]–[25].

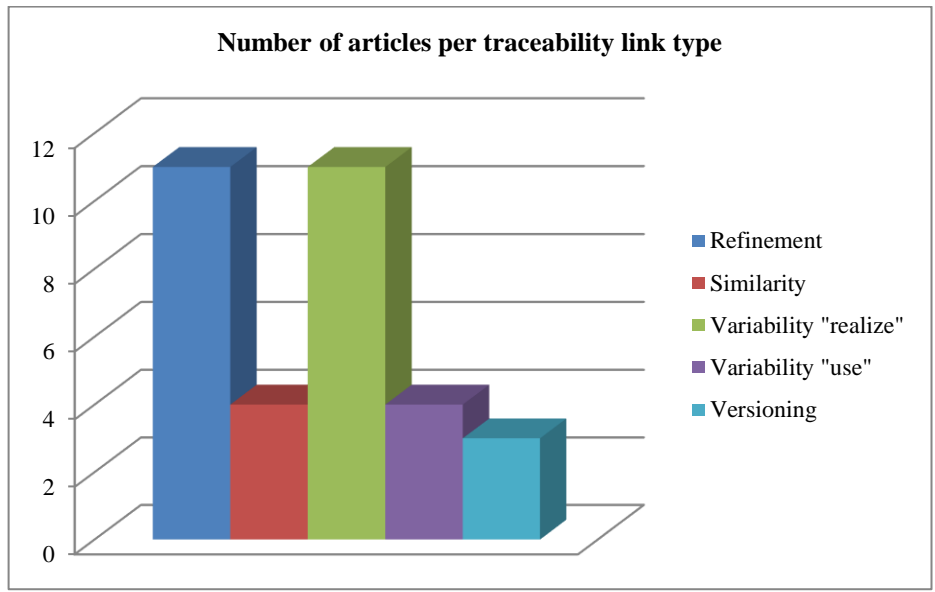

**Figure 6:** Number of articles per traceability link type

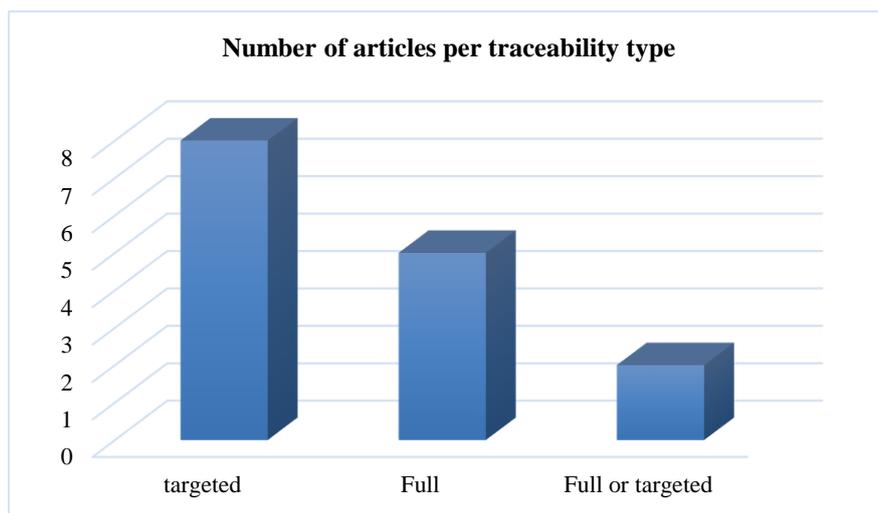

**Figure 7:** Number of articles per traceability type

**RQ4: What is the proposed solution to implement traceability in the Software Product Line?**

Graphic figure 8 presents the traceability solutions proposed according to their type. We distinguish 7 types of solutions: Approach, methodology, metamodel, model, algorithm, framework and tool. The traceability solutions proposed in the literature are mainly approaches. Indeed, 7 articles out of 15 propose approaches. For some, the approach is proposed alone, as for the article [20] which proposes to target the links to be drawn in a Software Product Line according to the objective behind the implementation of a traceability approach (evolution and maintenance, evaluation, ...). For this, the article defines 3 possible traceability paths: (i) the path between the features model and the corresponding structural model (implementation phase), (ii) the path between the Product Line level and the Product level and (iii) the path between the generic structural model (e.g.

patterns) and the concrete structural model (i.e. its implementation at the product level). One way or the other is to be chosen at the time of the implementation of the traceability to trace only the necessary links depending on the objective behind tracing. The article also emphasizes that multiple paths can coexist between two artifacts. Traces between two elements serving the same objectives are grouped into sets.

The article [21] proposes an approach accompanied by a methodology for refactoring products representing variation points in Software Product Lines. The proposed methodology allows the extraction of the feature - implementation and feature - feature traceability links, to then build a dependency graph, extract the variability information, and build the Product Line.

Article [22] deals with evolution of Software Product Lines and the tasks that the developer should perform to drive this evolution, while taking into account the various factors that can intervene and be impacted, in particular the consistency of traceability, the variability in terms of impacts on other products, and the accessibility of the system. For this, the article proposes an approach as well as a framework allowing its implementation. The proposed approach enables online updating of the Software Product Line through the use of viewpoints, an architectural solution that consists of a set of established conventional elements to define how to construct, interpret and use architectural views of a given system.

Four other papers, [12], [14], [16], [25], propose approaches accompanied by the implementation of a tool. The article [12] presents an approach that allows tracing between documents generated during the development of the Product Line. This approach defines rules for identifying traceability links and classifying them according to 9 different types of traceability: Satisfiability, dependency, overlaps, evolution, implements, refinement, containment, similar and different. These rules are expressed in a specific language, XQuery [26], while the documents are represented in XML. The article also presents a prototype for the implementation of this solution. The tool makes it possible to translate the documents into XML, to generate the traceability links from the predefined rules, as it also allows defining new rules.

In the article [14], the author proposes a use case-oriented approach to automate the maintenance and evolution of use cases in case of change in the configuration of the Software Product Line or/and its features, as it impacts the variability. It allows identifying the impact of a change (adding, deleting, updating...) before its implementation, and informing the analyst who initiated the change request. The solution is based on an iterative process until validation and application of the change decision. After this step, the changes are applied and a new use case diagram, retracing only the changes, is generated automatically. The proposed tool outputs, in addition to the specific product after reconfiguration, a report listing the changes made to the initial template. Also, the analyst is informed of the impacts identified and their possible causes.

The article [16] presents an approach that redefines Software Product Lines based on sets theory, from all features to all components, using traceability links. The relationships are coded as Quantified Boolean Formula (QBF). These are propositional logical formulas, with existential or universal quantifiers. They are represented as $Q_1x_1...Q_nx_n\ E(x_1...\ x_n)$, with E a propositional formula, $x_i$ propositional variables, and $Q_i$ existential or universal quantifier, for i between 1 and n [27]. These formulas are implemented by the proposed tool, called

Software Product Line ANalysis Engine (SPLANE).

The approach proposed by the article [25] aims to detect inconsistencies in variability between the specifications (features model) and the source code. Variability is modeled, at the implementation level, by a grouping of technical variability models (TVM) in the form of "forest structure", in contrast to the model of features that is rather modeled as "tree structure" ". The traceability links are formed from the mapping between features at the feature model level and their implementation at the TVM level. In order to establish this traceability, the proposed tool uses the DSL language: Domain Specific Language, written in Scala [28].

In addition, two articles present only a methodology [19], [23]. The first one, described above, and the second, for traceability at the level of variation points, based on the Unified Tabular Method mechanism to model the points of variation in Product Lines, which helps manage their complexity in terms of exponential number of possible scenarios. The method consists in a textual modeling, in a table, of the contents of the model of features as well as the information relating to the points of variation (variant, type, value (and therefore variables), subdomain, relation between the variables and the dependencies).

Four other articles propose frameworks [5], [15], [18], [22]. The frameworks described in [5] and [22] have already been discussed above. In [15], the motivation behind the proposed framework comes from the need to document the software product line architecture for a better understanding of the decisions made beforehand to facilitate and better guide future decision-making and system maintenance. The solution presented in this article proposes a model allowing better knowledge of the architecture of the Product Line and its variability. This model, called Product Line Architecture Knowledge (PLAK), combined with the Design Decision concept, and thanks to the use of traceability links, makes it possible to identify all the variables impacted by a modification decision in the Software Product Line architecture. This solution is proposed via a framework called Flexible Product Line Architecture (FPLA).

In [18], the author defines a framework for a Model Driven Software Product Line Engineering solution. The challenge in this type of solution is to keep a certain consistency of the data between all the models. The article proposes a transformation propagation chain between specifications, meta-model, model, and trace links. The proposed solution allows also identifying the possible changes and dependencies, always within the context of a model-oriented development.

Still in the context of "abstract and generic" solutions, and by grouping models and meta-models, we retain 3 solutions, those presented in the articles [6], [15], [17]. The study in [6] proposes a meta-model under UML that represents the interactions between the elements of the Product Line so that variability and traceability can be managed through these elements. This meta-model consists of two main parts: a part for product line management that includes project management and risk management, and a part representing the core of Product Line development, containing scoping, requirements and tests.

In addition, we have 5 concrete solutions presented in the form of tools as described above for the articles [12], [14], [16], [25], or as an algorithms like for the article [24].

In what follows we discuss the automation of the proposed solutions.

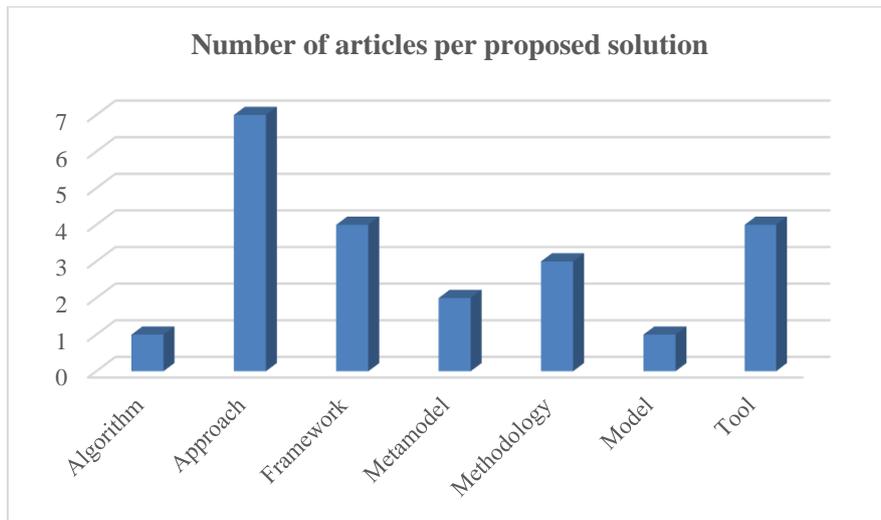

**Figure 8**: Number of articles per proposed solution

**RQ5: Is it a manual or automatic traceability?**

Through this research question, we discuss whether the Software Product Lines traceability solutions proposed in the literature are automation oriented. Indeed, it is clear that in a system as large and complex as Product Lines, maintaining traceability links while ensuring their integrity is a difficult task and its manual handling is error prone [2], hence the need for automatic traceability.

Figure 9 below represents the distribution of the articles previously selected, depending on whether the proposed traceability solution is manual, automatic or semi-automatic.

The graph shows that the trend is towards the automation of solutions (if we count the semi-automatic solutions also, proposed in articles [15] and [17]). This automation is present in the solutions proposed by the articles [5], [12], [14], [16], [18], [22], [25].

We note that, except for the article [6] that offers a manual solution for full traceability, all full traceability solutions are automatic [5], [12], [15]–[18]. This supports what is advanced in the literature, namely that automation minimizes the costs of adopting a traceability approach in the context of setting up a Software Product Line [29]–[31].

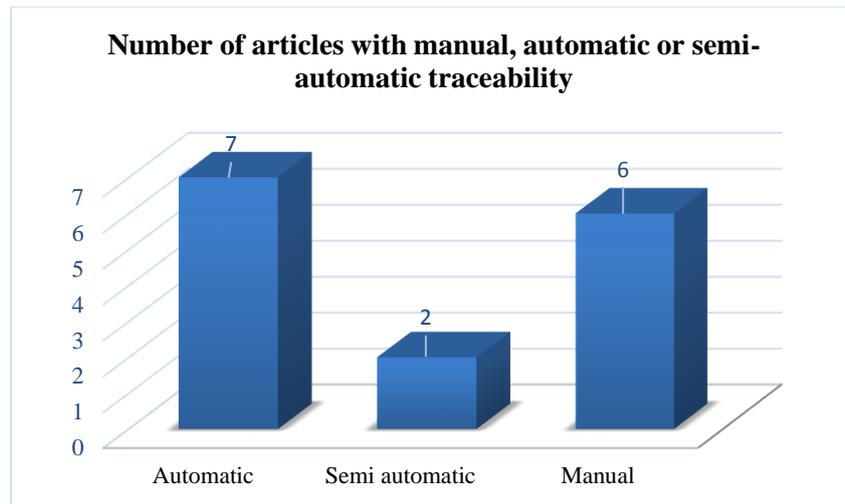

**Figure 9**: Number of articles with manual, automatic or semi-automatic traceability

**4. Constraints and limitations of the Review**

While conducting our systematic review of tracing solutions in Software Product Lines, we tried to be as impartial and objective as possible. We also tried to scan a maximum of studies in order to get significant and representative results. However, we encountered some difficulties and limitations. First, we could not browse all existing and possible sources: Some of them are difficult to exploit, like Google Scholar, as it does not propose an export option for the search results, and some might propose redundant information with other sources (Google Scholar, Researchgate …). Therefore, even if we tried to select the most relevant sources that include important studies, we could have missed some interesting articles for our research work.

Another difficulty that we encountered is the adaptation of search expression to the search source. In fact, each search source has its own logic, so we had to adapt the search expression, which is not that easy. Also, we are aware that some articles use keywords that might not be taken into consideration. We did our best to use the correct and the most generic search expression, but the risk of missing studies exists.

**5. Conclusion**

We presented through this article our systematic review for tracing solutions in Software Product Lines. 4 databases were concerned: ACM, IEEE Xplore, Elsevier, and Springer link; and we tried to answer 5 research questions related to tracing solutions in Software Product Lines. As a result of a first selection from the concerned databases, we get 9,449 articles to which we applied several predefined filters. We ended up with 15 articles that we analyzed according to our research questions.

As a result to our analysis, we can list several findings:

- Few research studies focus on the problem of defining a traceability solution in Software Product Lines.
- Traceability in Software Product Lines is implemented mainly in the Domain Engineering phase, followed by

the maintenance phase.

- The traceability links of refinement and variability types are the most used ones in the literature. Versioning type traceability is the least used.
- In the literature, researchers are more interested in modeling and architecture solutions, rather than versioning, even if the latter remains as important for the integrity of the proposed solution.
- Full traceability solutions focus on end-to-end traceability from features to implementation. They are part of the articles presenting links of traceability refinement type.
- Articles offering targeted traceability deal mainly with solutions relating to variability points.
- Modeling solutions are the most proposed in the literature, few studies offer more concrete solutions for implementing traceability in Software Product Lines.
- Solutions offering full traceability are automated.

As a reminder, this review was conducted on studies published between 2008 and 2018.

**6. Recommendations**

Literature is still in need of proposals for traceability solutions in the context of Software Product Lines. These solutions have to be concrete, targeted and costless so they can attract all stakeholders, including engineers and managers.

**References**

[1] Center of Excellence for Software Traceability, "What is Traceability? | Traceability." [Online]. Available: http://www.coest.org/index.php/what-is-traceability. [Accessed: 06-Jun-2014].

[2] J. Cleland-Huang, O. Gotel, and A. Zisman, Eds., *Software and Systems Traceability*. London: Springer London, 2012.

[3] J. Cleland-Huang, O. Gotel, J. H. Hayes, P. Mäder, and A. Zisman, "Software Traceability: Trends and Future Directions," in *36th International Conference on Software Engineering (ICSE),* 2014.

[4] M. Lindvall and K. Sandahl, "Practical Implications of Traceability," *Softw. Pract. Exp.*, vol. 26, no. 10, pp. 1161–1180, Oct. 1996.

[5] N. Anquetil, U. Kulesza, R. Mitschke, A. Moreira, J.-C. Royer, A. Rummler, and A. Sousa, "A model-driven traceability framework for software product lines," *Softw. Syst. Model.*, vol. 9, pp. 427–451, 2010.

[6] Y. C. Cavalcanti, I. do Carmo Machado, P. A. da Mota, S. Neto, L. L. Lobato, E. S. de Almeida, and S. R. de Lemos Meira, "Towards metamodel support for variability and traceability in software product lines," in *Proceedings of the 5th Workshop on Variability Modeling of Software-Intensive Systems - VaMoS '11*, 2011, pp. 49–57.

[7] B. Ramesh and M. Jarke, "Towards Reference Models for Requirements Traceability," *Softw. Eng. IEEE Trans.*, vol. 27, no. 1, pp. 58–93, 2001.

[8]  G. Spanoudakis and A. Zisman, "Software Traceability: A Roadmap," in *Handbook Of Software Engineering And Knowledge Engineering*, vol. III, WORLD SCIENTIFIC, 2005, pp. 395–428.

[9]  W. Jirapanthong and A. Zisman, "Supporting product line development through traceability," in *Proceedings - Asia-Pacific Software Engineering Conference, APSEC*, 2005, vol. 2005, pp. 506–514.

[10]  N. Anquetil, B. Grammel, I. Galvão, J. Noppen, S. Khan, H. Arboleda, A. Rashid, and A. Garcia, "Traceability for Model Driven, Software Product Line Engineering," in *ECMDA Traceability Workshop Proceedings*, 2008, vol. 12, pp. 77–86.

[11]  K. Berg, J. Bishop, and D. Muthig, "Tracing Software Product Line Variability - From Problem to Solution Space," *Proc. 2005 Annu. Res. Conf. South Africain Inst. Comput. Sci. Inf. Technol. IT Res. Dev. Ctries.*, pp. 182–191, 2005.

[12]  W. Jirapanthong and A. Zisman, "XTraQue: traceability for product line systems," *Softw. Syst. Model.*, vol. 8, no. 1, pp. 117–144, Feb. 2009.

[13]  B. Kitchenham, "Procedures for Performing Systematic Reviews," Keele, UK, 2004.

[14]  I. Hajri, A. Goknil, L. C. Briand, and T. Stephany, "Change impact analysis for evolving configuration decisions in product line use case models," *J. Syst. Softw.*, vol. 139, pp. 211–237, 2018.

[15]  J. Diaz, J. Perez, J. Garbajosa, and C. Fernandez-Sanchez, "Modeling product-line architectural knowledge," *Proc. Annu. Hawaii Int. Conf. Syst. Sci.*, vol. 2015-March, pp. 5383–5392, 2015.

[16]  S. Mohalik, S. Ramesh, J.-V. Millo, S. N. Krishna, and G. K. Narwane, "Tracing SPLs precisely and efficiently," in *Proceedings of the 16th International Software Product Line Conference on - SPLC '12 - volume 1*, 2012, pp. 186–195.

[17]  A. Espinoza, G. Botterweck, and J. Garbajosa, "A formal approach to reuse successful traceability practices in SPL projects," in *Proceedings of the 2010 ACM Symposium on Applied Computing - SAC '10*, 2010, pp. 2352–2359.

[18]  C. K. F. Corrêa, "Towards automatic consistency preservation for model-driven software product lines," *Proc. 15th Int. Softw. Prod. Line Conf. - SPLC '11*, p. 1, 2011.

[19]  S. H. Ripon, "A Unified Tabular Method for Modeling Variants of Software Product Line," *ACM SIGSOFT Softw. Eng. Notes*, vol. 37, no. 3, pp. 1–7, 2012.

[20]  P. Lago, H. Muccini, and H. van Vliet, "A scoped approach to traceability management," *J. Syst. Softw.*, vol. 82, no. 1, pp. 168–182, 2009.

[21]  L. Linsbauer, R. E. Lopez-Herrejon, and A. Egyed, "Variability extraction and modeling for product variants," *Softw. Syst. Model.*, vol. 16, no. 4, pp. 1179–1199, 2017.

[22]  D. Weyns, B. Michalik, A. Helleboogh, and N. Boucke, "An Architectural Approach to Support Online Updates of Software Product Lines," *2011 Ninth Work. IEEE/IFIP Conf. Softw. Archit.*, pp. 204–213, 2011.

[23]  H. Eyal-Salman, A. D. Seriai, and C. Dony, "Feature-to-code traceability in a collection of software variants: Combining formal concept analysis and information retrieval," *Proc. 2013 IEEE 14th Int.*

*Conf. Inf. Reuse Integr. IEEE IRI 2013*, pp. 209–216, 2013.

[24] L. Linsbauer, E. R. Lopez-Herrejon, and A. Egyed, "Recovering traceability between features and code in product variants," *Proc. 17th Int. Softw. Prod. Line Conf. - SPLC '13*, p. 131, 2013.

[25] X. Tërnava and P. Collet, "Early Consistency Checking between Specification and Implementation Variabilities," *Proc. 21st Int. Syst. Softw. Prod. Line Conf. - Vol. A - SPLC '17*, pp. 29–38, 2017.

[26] "XQuery." [Online]. Available: https://www.w3.org/TR/xquery/all/.

[27] M. Cadoli, A. Giovanardi, and M. Schaerf, "An Algorithm to Evaluate Quantified {Boolean} Formulae," in *AAAI/IAAI*, 1998, pp. 262–267.

[28] "DSL-Scala." [Online]. Available: https://github.com/ternava/variability-cchecking.

[29] H. U. Asuncion, F. François, and R. N. Taylor, "An end-to-end industrial software traceability tool," in *Proceedings of the the 6th joint meeting of the European software engineering conference and the ACM SIGSOFT symposium on The foundations of software engineering - ESEC-FSE '07*, 2007, pp. 115–124.

[30] A. Egyed, "A scenario-driven approach to trace dependency analysis," *IEEE Trans. Softw. Eng.*, vol. 29, no. 2, pp. 116–132, 2003.

[31] J. Cleland-Huang, R. Settimi, O. BenKhadra, E. Berezhanskaya, and S. Christina, "Goal-centric traceability for managing non-functional requirements," *Proc. 27th Int. Conf. Softw. Eng. - ICSE '05*, p. 362, 2005.